\documentclass[conference]{IEEEtran}
\IEEEoverridecommandlockouts
\usepackage{cite}
\usepackage{amsmath,amssymb,amsfonts}
\usepackage{algorithmic}
\usepackage{graphicx}
\usepackage{textcomp}
\usepackage{xcolor}
\usepackage{subcaption}
\usepackage{booktabs}
\usepackage{url}
\def\BibTeX{{\rm B\kern-.05em{\sc i\kern-.025em b}\kern-.08em
    T\kern-.1667em\lower.7ex\hbox{E}\kern-.125emX}}

\begin{document}

\title{Exploring the Distinctive Tweeting Patterns of
Toxic Twitter Users}

\author{
  \IEEEauthorblockN{
    \begin{tabular}{c}
      1\textsuperscript{st} Hina Qayyum \\
      \textit{Macquarie University} \\
      Sydney, Australia \\
      hina.qayyum@students.mq.edu.au
    \end{tabular}
    \hspace{2em}
    \begin{tabular}{c}
      2\textsuperscript{nd} Muhammad Ikram \\
      \textit{Macquarie University} \\
      Sydney, Australia \\
      muhammad.ikram@mq.edu.au
    \end{tabular}
    \hspace{2em}
    \begin{tabular}{c}
      3\textsuperscript{rd} Benjamin Zi Hao Zhao \\
      \textit{Macquarie University} \\
      Sydney, Australia \\
      ben\_zi.zhao@mq.edu.au
    \end{tabular}
  }

  \vspace{1.5em}

  \IEEEauthorblockN{
    \begin{tabular}{c}
      4\textsuperscript{th} Ian D. Wood \\
      \textit{Macquarie University} \\
      Sydney, Australia \\
      ian.wood@mq.edu.au
    \end{tabular}
    \hspace{2em}
    \begin{tabular}{c}
      5\textsuperscript{th} Nicolas Kourtellis \\
      \textit{Telefonica Research} \\
      Barcelona, Spain \\
      nicolas.kourtellis@telefonica.com
    \end{tabular}
    \hspace{2em}
    \begin{tabular}{c}
      6\textsuperscript{th} Mohamed Ali Kaafar \\
      \textit{Macquarie University} \\
      Sydney, Australia \\
      dali.kaafar@mq.edu.au
    \end{tabular}
  }
}

\maketitle

\begin{abstract}

In the pursuit of bolstering user safety, social media platforms deploy active moderation strategies, including content removal and user suspension. These measures target users engaged in discussions marked by hate speech or toxicity, often linked to specific keywords or hashtags. Nonetheless, the increasing prevalence of toxicity indicates that certain users adeptly circumvent these measures.

This study examines consistently toxic users on Twitter (rebranded as $\mathbb{X}$) Rather than relying on traditional methods based on specific topics or hashtags, we employ a novel approach based on patterns of toxic tweets, yielding deeper insights into their behavior.

We analyzed 38 million tweets from the timelines of 12,148 Twitter users and identified the top 1,457 users who consistently exhibit toxic behavior, relying on metrics like the Gini index and Toxicity score. By comparing their posting patterns to those of non-consistently toxic users, we have uncovered distinctive temporal patterns, including contiguous activity spans, inter-tweet intervals (referred to as ``Burstiness''), and churn analysis. These findings provide strong evidence for the existence of a unique tweeting pattern associated with toxic behavior on Twitter.

Crucially, our methodology transcends Twitter and can be adapted to various social media platforms, facilitating the identification of consistently toxic users based on their posting behavior. This research contributes to ongoing efforts to combat online toxicity and offers insights for refining moderation strategies in the digital realm. We are committed to open research and will provide our code and data to the research community.
\end{abstract}


\maketitle
\section{Introduction}
With the primary goal of ensuring the safety and well-being of its users, Twitter (Note: Even though Twitter has undergone a rebranding to become $\mathbb{X}$, the dataset employed in this study was collected from the platform during the period when it retained the name Twitter. Thus, we consistently use the term ``Twitter" throughout this paper.) implements a range of measures to uphold its terms of service. One of these vital actions involves actively moderating the platform, which includes the removal of inappropriate or harmful content and even the suspension of certain users. These moderation efforts are typically triggered when users engage in conversations that promote hatred, often identified through the use of specific keywords or hashtags.

Despite these proactive efforts to maintain a positive online environment, there still remains a challenge in dealing with toxic users who manage to evade these moderating techniques. Therefore, it may prove advantageous to delve deeper into the users and behaviors of these toxic users. Exploring often overlooked aspects of these users could potentially shed light on their activity and posting patterns, leading to more effective and targeted measures in curbing their negative impact on the platform and its community.

Several recent research papers have examined the benefits of content moderation methods and their impact on reducing the overall toxicity within the platform, particularly when considering various subjects or characteristics associated with user users. Such as~\cite{hanley2023twits} analyzed the correlation between various account-level attributes, such as political ideology and account age, and their impact on the frequency of posting toxic content by Twitter users.~\cite{maj2020a} According to their findings, approximately 21\% of the entire discussion related to the COVID-19 pandemic and the WHO's involvement in managing the crisis comprised toxic messages. However, ~\cite{deverna2022identification} pointed out that the existing moderation techniques are failing to address a substantial portion of consistently toxic and misinformation-spreading political users. 

This research paper aims to conduct a pilot study that examines the tweeting patterns of users on the Twitter platform who consistently share toxic content. We hypothesize that these toxic Twitter users exhibit a specific tweeting pattern that significantly contrasts with users who only occasionally post toxic content. To the best of our knowledge, no prior research has specifically examined consistently toxic users regardless of the topics they post about. The continuous escalation of toxicity on Twitter can be attributed to the prolonged existence of such accounts.

We look into the longitudinal data or timelines of the 12K Twitter users, and based on toxicity scores of their 3,200 tweets per user, we identify that 12\% (1,457) users were consistently posting toxic tweets. We compare the tweeting pattern of these users with the rest of the 88\% (10,727) users. 
We explore different aspects of tweeting patterns that are associated with the consistent toxic behavior of users on Twitter. Our analysis encompasses the duration of the tweeting activity, the distribution of time intervals between consecutive tweets referred to as Burstiness, and an evaluation of churn patterns of overall activity. By examining these factors, we gain a comprehensive understanding of how these users engage with the platform and their tendencies in terms of posting frequency and account continuity. All these statistical measures show that consistently toxic users do indeed have a distinguished tweeting pattern. The main contributions of this paper are listed below:
\begin{itemize}
    \item A first study to explore the tweeting pattern of Twitter users 
    with a methodology that is independent of the platform and can be extended to toxic users on any platform.
    \item We propose the incorporation of straightforward statistical metrics, namely the Gini coefficient, Burstiness, and churn model, for evaluating the activities and posting behaviors of Twitter users and identifying the consistently toxic users. 
    \item Upon publication, we openly release a longitudinal dataset comprising 12,148 users, containing tweet IDs of consecutive timeline tweet timestamps, toxicity scores associated with tweets, and supplementary experimental code.
\end{itemize}
Section~\ref{sec:related-work} includes the comprehensive literature review, while Section~\ref{sec:dataset and methods} encompasses the details of dataset collection (Section~\ref{sec:dataset}), and further subsections (Section~\ref{sec:Toxicity scores}, Section~\ref{sec:Gini index of toxicity scores}, and Section~\ref{sec:Consistently toxic users}) expound on the process of selecting consistently toxic users. Subsections (Section~\ref{sec:Activity span}, Section~\ref{sec:Bot scores}, Section~\ref{sec:Burstiness}), and Section~\ref{sec:Churn analysis} delve into the methods employed. Section~\ref{sec:Results} elaborates on all deduced results, while the final Section~\ref{sec:discussion and conclusion} encompasses the implications of these results and the concluding remarks.

\section{Literature review}
\label{sec:related-work}


In the context of our unprecedented study focusing on consistently toxic Twitter users, it is paramount to conduct a comprehensive review of the existing body of literature. This literature review serves as the foundation for our research, offering insights into previous studies, their methodologies, and critical gaps that our study aims to address. We begin by providing an overview of the background and context of content moderation and then delve into the critical analysis of previous research, identifying strengths, weaknesses, and areas for further exploration.


Content moderation is a governance method that orchestrates the involvement of individuals within an online community to promote collaboration and deter misuse~\cite{grimmelmann2015virtues}. It establishes criteria for retaining online posts and users, determines deletions or temporary deactivations, and influences the visibility of approved posts. Social media platforms formulate content rules within terms of service and guidelines that users encounter during registration. These often intricate rules are simplified as ``community guidelines" to clarify banned content and prohibited actions, including child exploitation, terrorism, pornography, spam, malware, and phishing~\cite{cao2015detecting, singhal2019analysis, singhal2023cybersecurity, zhang2012detecting}.

While community guidelines serve as a foundation for content moderation, they face challenges related to clarity and enforcement. The complexity of these guidelines is a potential barrier to user understanding and compliance~\cite{steinfeld2016agree}. Moreover, the effectiveness of content moderation depends on how well platforms strike a balance between protecting free expression and curbing harmful content. Addressing these challenges is critical for maintaining online communities that are both safe and inclusive.


Terms of service and community guidelines outline actions taken by platforms when detecting service violations or abusive behavior, ranging from individual post moderation to account suspension~\cite{le2021setting, myers2018censored}.

Hard moderation represents the strictest method through which a platform enforces its policies, involving the removal of content or entities from the platform~\cite{chandrasekharan2022quarantined, singhal2023sok, saleem2018aftermath}.

Soft moderation has become the initial strategy for platforms to address content violating their guidelines. In this approach, content is not removed but users are alerted about potential issues through warning labels, additional context labels, or limited reach, achieved by quarantining the uncertain content ~\cite{geeng2020social, mena2020cleaning, zannettou2021won}. Recent research has demonstrated that moderation techniques do reduce the prevalence of toxic topic discussions and overall platform toxicity~\cite{10.1145/3479525}.


While moderation mechanisms play a vital role in content control, questions arise regarding their scalability and adaptability. The effectiveness of hard and soft moderation strategies depends on the platform's ability to keep pace with evolving forms of toxic behavior. Achieving a delicate balance between preserving free speech and curbing abuse is an ongoing challenge. It is essential to explore new approaches to content moderation that can effectively address emerging threats.


Recent surveys have assessed detection methods concerning hate speech or misinformation. Detection systems utilize text features from social media posts, including TF-IDF and Part of Speech (POS) tags. Lexicon-based techniques are common in hate speech detection, while misinformation detection relies on propagation structures, crowd intelligence, and knowledge-based methods~\cite{fortuna2018survey, guo2019future, islam2020deep, schmidt2017survey, yin2021towards, zhou2020survey}.

Recent advances in Deep Neural Networks (DNNs) have revolutionized both hate speech and misinformation detection, eliminating the need for extensive feature engineering and domain expertise. DNNs have been applied to detect hate speech and misinformation in various formats, including text, images, and videos~\cite{modha2019ld, wu2018tracing, pramanick2021momenta, wang2018eann}.


While significant progress has been made in detection techniques, several challenges persist. Lexicon-based approaches may struggle with context and evolving language use. Detecting misinformation in multimedia formats, including images and videos, presents unique challenges that require innovative solutions. Additionally, addressing biases present in training data is crucial for ensuring fair and effective detection. There is an ongoing need for more robust, adaptive, and unbiased detection techniques to combat the evolving landscape of toxic content.


Qayyum et al.~\cite{10.1145/3578503.3583619} address the challenge of characterizing toxic content by adopting a profile-centric approach. Their dataset is extensive, consisting of 293 million tweets, but they only look at the top toxic profiles in their dataset and do not take into account the consistency of toxic content over time.


Profile-centric approaches provide valuable insights into toxic users, yet the focus on top toxic profiles may overlook consistently toxic users who may not be highly prominent. Understanding the consistency of toxic content over time is a critical dimension, as it allows for a more comprehensive assessment of the scope and impact of toxicity on social media platforms.


Until now, no research has investigated the identification of toxic users and their hateful content using tweeting patterns. This study examines the tweeting patterns of consistently toxic users on Twitter. This approach offers a more intuitive perspective on moderating toxic or hateful content. These insights can stand alone or complement other detection methods, enhancing content moderation not only on Twitter but across various platforms~\cite{10.1145/3578503.3583619}.


Our study addresses a critical gap in existing research by focusing on consistently toxic users and their tweeting patterns. By delving into the temporal aspects of toxicity, we aim to provide a nuanced understanding of how these users operate. However, challenges in defining and identifying consistently toxic users persist, and the scalability of our approach to diverse platforms remains to be explored. Additionally, while our method contributes to content moderation efforts, it is not a panacea, and a holistic approach to addressing online toxicity is necessary.


The synthesis of the literature reveals that content moderation, toxic content detection, and characterization of toxic users are active areas of research. While progress has been made in identifying toxic content and users, there is still much to be explored in terms of consistently toxic users' tweeting patterns and their long-term impact on social media platforms.


The critical analysis of existing research highlights several research gaps and areas for future exploration. These include:
\begin{itemize}
\item Developing more robust and adaptive toxic content detection techniques, particularly for multimedia formats.
\item Investigating the scalability and generalizability of content moderation methods to diverse online platforms.
\item Exploring innovative approaches to characterizing consistently toxic users and assessing their long-term impact.
\item Collaborative efforts to address biases and ethical considerations in content moderation and detection.
\end{itemize}

In conclusion, this comprehensive literature review has provided insights into the existing body of research related to content moderation, toxic content detection, and the characterization of toxic users. Our study builds upon this foundation by focusing on the tweeting patterns of consistently toxic users on Twitter, contributing to a deeper understanding of online toxicity and content moderation strategies.

\section{dataset and methods}
\label{sec:dataset and methods}
\subsection{ Dataset}
\label{sec:dataset}
We utilize a Twitter dataset dedicated to studying toxicity and misconduct on the platform, as cited in \cite{10.1145/3578503.3583619}. This dataset offers valuable insights into the activity patterns and content of toxic Twitter users.

Out of the 143,000 users available in the dataset, only 12,184 users have a total of 3,200 tweets each. The remaining users have varying tweet counts, ranging from 7 to several hundred tweets. To ensure a substantial number of timeline events and enable fair comparisons of posting behaviors over time, we narrow our focus to users with exactly 3,200 tweets. This number represents the maximum upper limit set by the Twitter API.

Each Twitter user's timeline comprises tweets posted at specific times indicated by timestamps. These timelines capture tweet timestamps and tweet contents, including any embedded hashtags or URLs. In total, our dataset consists of 38,998,600 tweets, which is the result of multiplying 12,148 users by 3,200 tweets each.
\subsection{Toxicity scores}
\label{sec:Toxicity scores}
Due to the dataset's extensive size, we were unable to manually annotate the toxicity level of tweet text. Instead, we employed Google's Perspective API~\cite{perspective} to ascertain the ``Toxicity" score for the text within each user's tweet timeline. The Perspective API employs machine learning models to assign probability scores ranging from 0 to 1, indicating the extent of toxicity in the text. A higher probability score indicates a higher level of toxicity. In total, we queried the Perspective API to calculate the toxicity score for all 38 million tweets originating from 12,184 users.

\subsection{Gini Index}
\label{sec:Gini index of toxicity scores}
Subsequently, we utilize the Gini Index to assess the concentration of toxicity scores among users based on their tweet text. The Gini Index, originally devised to indicate wealth concentration~\cite{gini1912variabilita}, is employed as a metric in this context. A set of consistently low or high toxicity scores results in a value closer to 0, while a wide spectrum of both low and high scores yields a Gini Index value nearing 1. We compute and compare the Gini Indices of toxicity scores (based on 3,200 tweets) for all 12K users in our dataset. Figure~\ref{fig:gini_index_and_toxicity_scores_per_user.pdf} illustrates the cumulative distribution function (CDF) of mean toxicity scores and the Gini index of toxicity scores per user.

It's worth noting that within our dataset, approximately 95\% of users exhibit a mean tweet toxicity score under 0.3. Just \~5\% of Twitter users fall within the mean toxicity score range of 0.3 to 0.9. Conversely, approximately \~5\% of users possess a Gini index below 0.3, indicating consistent behavior in terms of toxicity or non-toxicity. Meanwhile, 90\% of users boast a Gini index ranging from 0.3 to 0.5, reflecting a mixture of both toxic and non-toxic scores.
\subsection{Consistently toxic users (CTUs)}
\label{sec:Consistently toxic users}
We are primarily focused on investigating whether Twitter users displaying consistent toxic behavior demonstrate unique patterns in their tweeting behavior. To address this, we aim to identify users who consistently post toxic tweets.

\begin{figure}[!ht]
    \centering
    \begin{subfigure}[b]{.75\columnwidth}
    \centering
    \includegraphics[width=\textwidth]{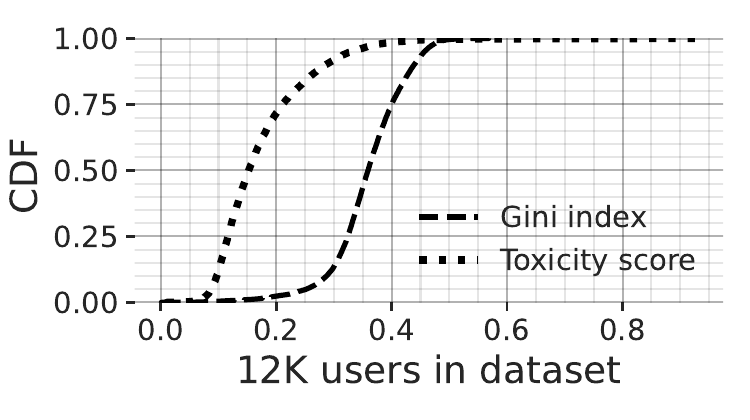}
    \caption{}\label{fig:gini_index_and_toxicity_scores_per_user.pdf}
    \end{subfigure}
    \begin{subfigure}[b]{.75\columnwidth}
    \centering
    \includegraphics[width=\textwidth]{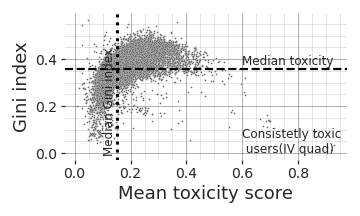}
    \caption{}\label{fig:scatter_mean_toxicity_and_gini.png}
    \end{subfigure}

    \vspace{-2mm}
     \caption{\small (\protect\subref{fig:gini_index_and_toxicity_scores_per_user.pdf}) A cumulative distribution function of mean toxicity score (dotted line) and Gini index of the toxicity scores per user (dashed line); (\subref{fig:scatter_mean_toxicity_and_gini.png}) A scatter plot of 12K users with Gini index on the y-axis and the mean toxicity scores on the y-axis. The horizontal dashed line represents the median Gini index calculated on all the users and a vertical dotted line is a median toxicity score.}
    \label{}
\end{figure}

The process is illustrated in Figure~\ref{fig:scatter_mean_toxicity_and_gini.png}. By calculating the mean toxicity scores and Gini index across all users, we establish reference values. Spearman's correlation between the mean toxicity and the Gini index stands at 0.673, while the p-value is 0, highlighting a significant inverse correlation between the Gini index and mean toxicity scores.

The median of the mean toxicity score registers at 0.15, while the median Gini index is 0.359. These medians provide insights into the behavior of an average user within our dataset. To identify consistently more toxic users, we focus on users falling equal to or below the median Gini index and equal to or above the median toxicity score (IV quadrants in Figure~\ref{fig:scatter_mean_toxicity_and_gini.png}). In total, 1,457 users consistently exhibit toxic behavior, referred to as ``Consistently Toxic Users" or ``CTUs" in our study. The remaining 10,727 users constitute our baseline set, labeled as ``Baseline Users'' or ``BUs'' for brevity. Thus, we define a toxic user as follows:

{\bf``A consistently toxic Twitter user (CTU) is a profile who regularly posts tweets that contain harmful or toxic content."}

In terms of methodology, opting for the mean toxicity of tweets per user allows us to mitigate the impact of occasional highly toxic tweets, possibly stemming from the Perspective API. This approach of combining the mean toxic score per user with the Gini index proved more sensible and yielded enhanced outcomes. The utilization of median dataset values to identify consistently toxic users was deliberate, and aimed at capturing outlier users effectively. A summary of our finalized dataset is outlined in Table~\ref{tab:dataset_overview}.

\begin{table}[h!]
\centering
\resizebox{0.90\columnwidth}{!}{
{\large
\begin{tabular}{r|r|r}
\toprule
{\bf }& {\bf \# Users}& {\bf \#Total tweets}\\

\midrule
{\bf Total users in dataset}& {\bf 12,184}&  {\bf 38,988,600} \\\hline
Consistently Toxic users ({\bf CTUs})&1,457&4,662,400\\\hline
Baseline users ({\bf BUs})&10,727&34,326,400\\
\bottomrule
\end{tabular}
}
}
\caption{\small Overview of the dataset; Consistently toxic users are referred to as CTUs and baseline as BUs for brevity (Section~\ref{sec:Consistently toxic users}).}
\vspace{-5mm}
\label{tab:dataset_overview}
\end{table}
\subsection{Activity span}
\label{sec:Activity span}
The Twitter API provided us with access to 3200 tweets for each user, systematically organized with chronological sequencing and accompanied by corresponding timestamps. These timestamps indicate the precise time of posting for each tweet and maintain their uniqueness within each user's timeline. The interval between the most recent and earliest tweet establishes the temporal extent of a Twitter user's  ``Activity span''. We evaluate this activity span in terms of total years in a user's timeline.

For instance, if a user consistently publishes an average of 200 tweets every day, their activity span on Twitter would amount to 16 days, analogous to a year. Conversely, a user who shares 200 tweets annually would have an activity span extending to 16 years. Individuals who participate in frequent tweeting will experience briefer activity spans, whereas those who tweet less often will observe extended gaps between their initial and latest tweets. We recognize that these timelines might not cover a user's complete posting history. Still, having access to 3,200 tweets provides a substantial dataset for inferring a user's customary posting patterns.
\subsection{Bot scores}
\label{sec:Bot scores}
The ``Botometer API v4''\cite{Sayyadiharikandeh_2020} employs five specialized classifiers to assess the resemblance of a Twitter user to an automated account or a bot. This classification involves analyzing user features such as the number of friends (accounts followed), social network structure, temporal activity (including tweets, likes, and retweets), tweet language, and sentiment. By considering these features, the Botometer API distinguishes between bot accounts and human accounts. The API provides overall bot scores within the range of [0, 1]. These scores are generated using either English features or Universal features (language-independent), and we report the overall universal feature scores, a greater score indicates an increased likelihood of a Twitter account being a bot. 
\subsection{Burstiness}
\label{sec:Burstiness}
``Burstiness''~\cite{Kim_2016} of a given event sequence can be described in its inter-event time distribution, 
where the inter-event time ``$t$" is defined as the time interval between any two consecutive events. The inter-event time distribution score is known as the ``Burstiness Score"
$
    B=\frac{\sigma-\mu}{\sigma+\mu}=\frac{r-1}{r+1}
    \label{eq:burstiness}
$
, where $r=\sigma/\mu$ is the
coefficient of variation and $\sigma,\mu$ denote the standard deviation and mean of inter-event times respectively. 
$B$ ranges between -1 and 1, whereby -1 describes a periodic time series, 0 is a random sequence, and 1 is an extremely bursty time series (as $\sigma\rightarrow\inf$ for finite $\mu$).
\subsection{Churn analysis}
\label{sec:Churn analysis}
\begin{figure}[!ht]
    \centering
    \includegraphics[width=0.90\columnwidth]{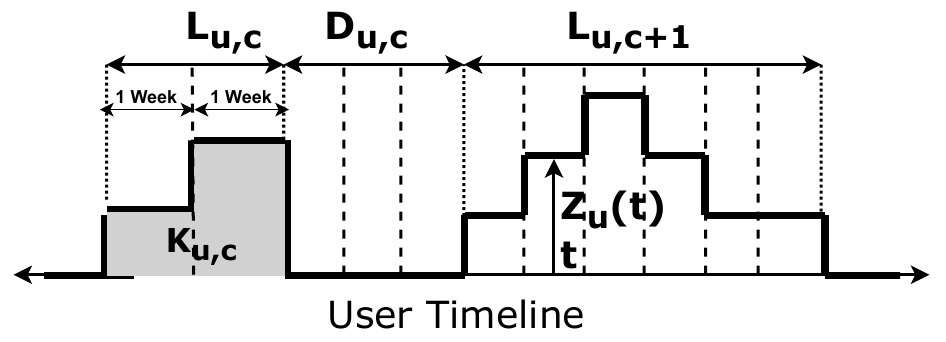}
    \vspace{-2mm}
     \caption{\small Churn Model. $L_{u,c}$ and $D_{u,c}$ are the life and deathtime (week granularity); $K_{u,c}$ is the sum of the toxicity scores of all tweets in one $L_{u}$.}
    \label{fig:churn_diagram.pdf}
\end{figure}
A ``Churn model'' delineates discrete temporal phases in which the Twitter users are active, characterized by consistent engagements like posting tweets, and inactive, signifying periods of diminished or absent participation. By quantifying the durations of these active and inactive phases, the churn model illuminates users' temporal engagement and withdrawal patterns, thereby contributing to the comprehension of their tweeting behavior dynamics.

We find inspiration from \cite{10.1145/3321705.3329834} to conceptualize the timeline of each user as an alternating renewal process of tweeting activity. This conceptual framework effectively breaks down the unique components that make up an individual's timeline within the context of a ``Churn model''. This model encapsulates their tweeting activity as well as periods of inactivity.

The churn pattern is captured through a personalized alternating renewal process denoted as $Z_u(t)$, specific to each user $u$, and based on the churn model described in~\cite{10.1145/3321705.3329834}. Within any given time frame, $Z_u (t) = 1$ indicates that user $u$ has posted a tweet at time $t$, while $Z_u (t) = 0$ signifies no activity. With $n$ users satisfying $1 \leq u \leq n$ and $t$ measured in weeks, we divide the complete timelines of users into weekly segments. This partition allows us to count both the total number of tweets and the cumulative toxicity scores for each time frame $K_{u,c}$, enhancing our understanding of toxicity during that specific period (a week).

The visual representation of the model can be observed in Figure~\ref{fig:churn_diagram.pdf}, where $c$ represents the specific cycle of user activity. Within each cycle, which can cover any number of weeks, there is either a constant phase of activity (tweeting) or a phase of inactivity (no tweeting). These periods are denoted as $L_{u,c} > 0$ for active phases and $D_{u,c} > 0$ for inactive phases. Constructing the churn model empirically for each user involves calculating an average lifespan (i.e., $L_{i,c=1\rightarrow\infty}$) and an average period of inactivity (i.e., $D_{i,c=1\rightarrow\infty}$) by taking averages across all cycles within the user's timeline. The average lifespan is symbolized as $L_u$, while the average period of inactivity is indicated as $D_u$.

\section{Results}
\label{sec:Results}
Given that our central hypothesis posits the presence of a unique tweeting pattern among toxic users, we proceed to present the outcomes of our investigation within this section.
\subsection{Tweeting behavior displayed by consistently toxic Twitter users (CTUs) diverges from that of automated bots.}
In the context of Twitter users, it is imperative to distinguish that consistently toxic users (CTUs) do not uniformly equate to automated bots. As depicted in Figure~\ref{fig:botometer_raw_scores_universal_overall.pdf}, an intriguing revelation surfaces. It becomes apparent that CTUs exhibit significantly lower Botometer scores in comparison to baseline users (BUs). This stark difference in scores indicates a notable trend among CTUs - their profiles tend to closely resemble human interaction patterns. In essence, these lower Botometer scores underscore the human-like nature of CTUs' online presence on the platform.

This finding carries implications for content moderation and the understanding of toxic behavior on Twitter. While automated bot accounts may indeed contribute to toxicity, it is essential to recognize that a substantial portion of consistently toxic behavior is exhibited by users with profiles that mirror typical human engagement. This nuanced understanding underscores the complexity of addressing toxicity on social media platforms.
\subsection{Consistently toxic users (CTUs) display unique tweeting patterns with intermittent or sporadic activity.}
\begin{figure}[t]
    \begin{subfigure}[b]{.24\textwidth}
    \centering
    \includegraphics[width=\textwidth]{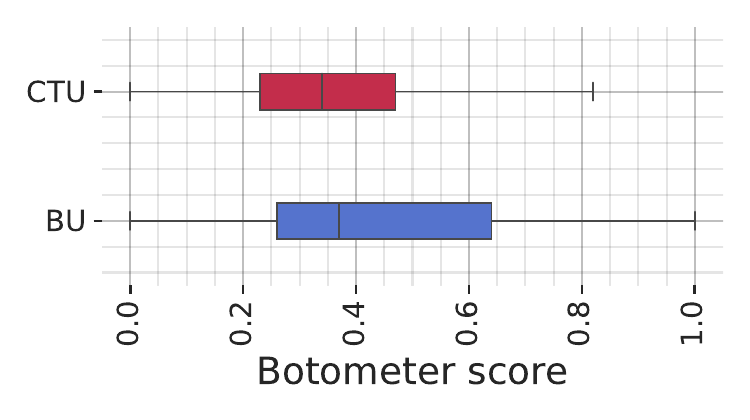}
    \caption{}\label{fig:botometer_raw_scores_universal_overall.pdf}
    \end{subfigure}
    \begin{subfigure}[b]{.24\textwidth}
    \centering
    \includegraphics[width=\textwidth]{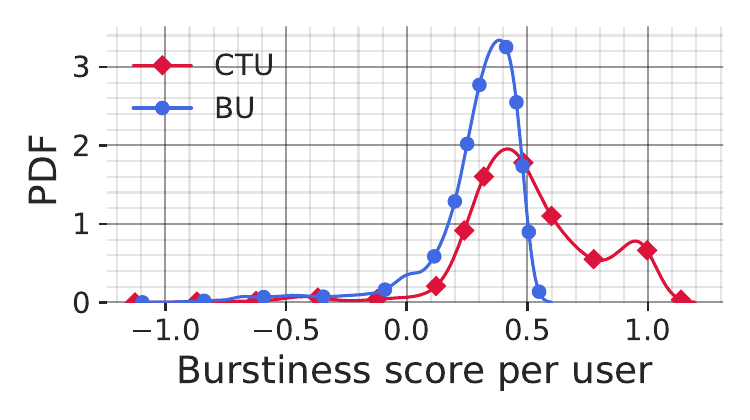}
    \caption{}\label{fig:PDF_burstiness_CTU_BU.pdf}
    \end{subfigure}
    
    \begin{subfigure}[b]{.24\textwidth}
    \centering
    \includegraphics[width=\textwidth]{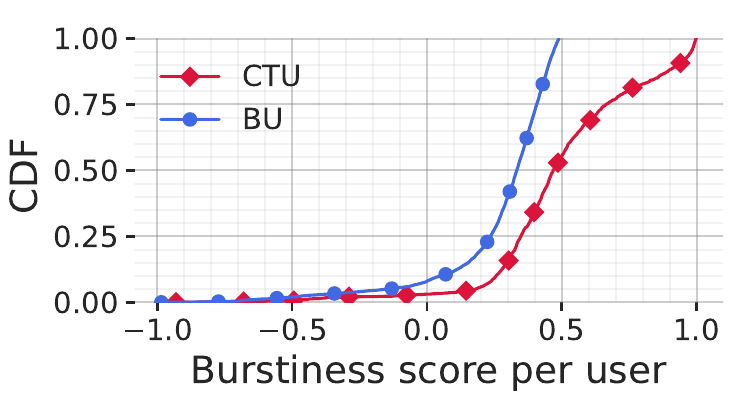}
    \caption{}\label{fig:CDF_burstiness_on_all_tweets.pdf}
    \end{subfigure}
    \begin{subfigure}[b]{.24\textwidth}
    \centering
    \includegraphics[width=\textwidth]{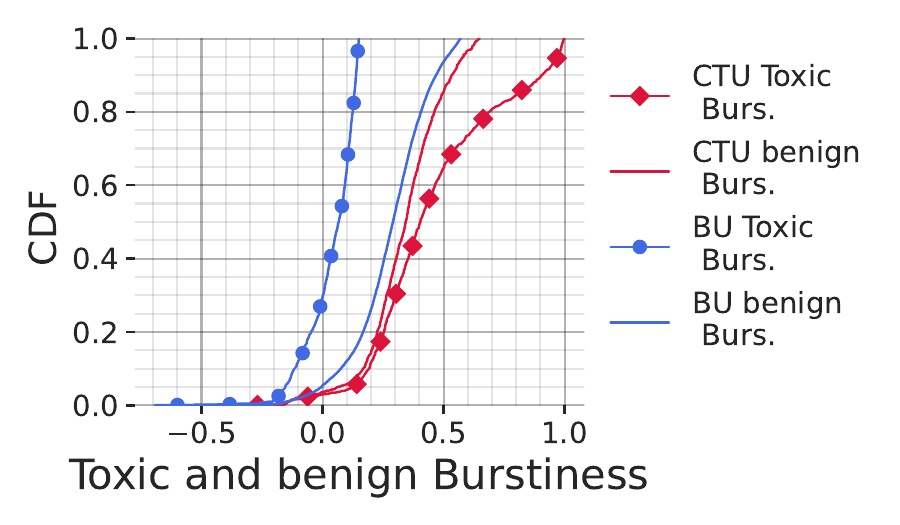}
    \caption{}\label{fig:toxic_nontoxic_burstiness.pdf}
    \end{subfigure}

     \caption{\small (\subref{fig:botometer_raw_scores_universal_overall.pdf}) Botometer scores of users in CTU and BUs; (\subref{fig:PDF_burstiness_CTU_BU.pdf}) Probability Distribution Function CDF of the Burstiness scores on all tweets of users in CTUs and BUs; (\subref{fig:CDF_burstiness_on_all_tweets.pdf}) Cumulative Distribution Function CDF of the Burstiness scores on all tweets of users in CTUs and BUs; (\subref{fig:toxic_nontoxic_burstiness.pdf}) Cumulative Distribution Function CDF of the Burstiness score calculated with the toxic and benign tweets per user.}
    \label{}
\end{figure}

In our analysis, we delve into the Burstiness score, a crucial metric (Section~\ref{sec:Burstiness}) that sheds light on the temporal posting patterns of Twitter users within the context of both consistently toxic users (CTUs) and baseline users (BUs). Burstiness is calculated by examining the entire collection of tweets across all users in both CTUs and BUs. The calculation is visually represented as a probability density function in Figure~\ref{fig:PDF_burstiness_CTU_BU.pdf}, with the resulting cumulative distribution function of Burstiness scores depicted in Figure~\ref{fig:CDF_burstiness_on_all_tweets.pdf}.

The data reveals a striking observation: CTUs tend to exhibit a more pronounced bursty posting pattern when compared to baseline users. This burstiness suggests that CTUs engage in tweet activity characterized by intermittent bursts of intense posting, followed by periods of lower activity. Notably, a significant subset of approximately 25\% of CTUs stands out, showcasing exceptionally high Burstiness scores ranging between 0.8 and 1.00. These scores indicate a substantial intensity of burst activity in their tweeting behavior.

To gain deeper insights, our investigation delves into potential disparities between the posting patterns of toxic and non-toxic tweets within both CTUs and BUs. Each tweet in a user's timeline receives a probabilistic toxicity score, as discussed in Section \ref{sec:Toxicity scores}. These scores range from 0 to 1, with higher values signifying increased toxicity. The mean toxicity score, detailed in Section~\ref{sec:Gini index of toxicity scores}, serves as a reference point.
We segment the 3,200 tweets per user into two categories: tweets with toxicity scores exceeding the mean are labeled as toxic, while those with scores below the mean are deemed benign. This segmentation allows us to explore how toxic and non-toxic tweets contribute to the burstiness of user posting patterns.

For each user in both CTU and BU categories, we calculate distinct Burstiness scores for their toxic and benign tweets. Consequently, each user in our dataset has two Burstiness scores—one characterizing the posting pattern of toxic tweets and the other indicating the burstiness of benign tweets. The cumulative distribution functions of these scores are depicted in Figure~\ref{fig:toxic_nontoxic_burstiness.pdf}.

These graphical representations provide valuable insights. It becomes evident that CTUs tend to concentrate their most toxic tweets in close temporal proximity, as reflected in Burstiness scores spanning from 0.25 to 1. This observation implies that CTUs engage in intense bursts of toxic content dissemination, contributing to the pronounced burstiness in their posting patterns. On the contrary, approximately 95\% of BUs exhibit a more random pattern in posting toxic tweets, characterized by Burstiness scores ranging from -0.25 to 0.15.

This distinction in posting behavior between CTUs and BUs, both in terms of burstiness and the distribution of toxic and non-toxic content, has significant implications. It not only deepens our understanding of how consistently toxic users engage with the platform but also provides valuable insights for content moderation strategies. Recognizing these patterns allows for more targeted approaches in addressing toxicity on Twitter, ultimately contributing to a safer and more constructive online environment.
\subsection{Toxic users (CTUs) display a reduced annual tweeting rate.}
\label{sec:Activity span annual}
In examining the annual tweeting activity (Section~\ref{sec:Activity span}), our investigation reveals a striking divergence between consistently toxic users (CTUs) and baseline users (BUs). Notably, baseline users tend to exhibit more prolific yearly tweeting frequencies compared to CTUs. A comprehensive breakdown of the tweeting activity patterns for both CTUs and BUs is presented in Table~\ref{tab:total_activity_span}. Approximately 70\% of BUs are actively engaged in posting tweets over a two-year window, with an additional 18\% reaching the maximum tweet count of 3200 within five years or less.

Of particular significance is the diminished annual tweeting rate observed among CTUs, who are consistently associated with toxic behavior. Intriguingly, around 35\% of CTUs persist in their toxic tweeting activity for over a decade. This finding underscores the enduring nature of toxicity in certain online user profiles, warranting further exploration into the factors contributing to this long-term toxic engagement.
\begin{table}[t!]
\centering
\resizebox{0.85\columnwidth}{!}{
{\large
\begin{tabular}{l|r|r|r }
\toprule
{\bf activity span }& {\bf \#toxic users}& {\bf \#baseline users}& {\bf \% of tws/year} \\
\midrule
1 year& 159(0.10\%)& 1,249(11.64\%)& 3,200\\\hline
2 year& 97(6.6\%)& 7,509(70.00\%)& 1,600\\\hline
3 year& 316(21.6\%)& 1,153(10.72\%)& 1,067\\\hline
4 year& 169(11.59\%)& 589(5.5\%)& 800\\\hline
5 year& 108(7.41\%)& 227(2.12\%)& 640\\\hline
6 year& 50(3.43\%)& 0& 533\\\hline
7 year& 26(1.78\%)& 0& 457\\\hline
8 year& 17(1.16\%)& 0& 400\\\hline
9 year& 8(0.54\%)& 0& 355\\\hline
10 year& 5(0.34\%)& 0& 320\\\hline
11 year& 2(0.13\%)& 0& 291\\\hline
12 year& 116(7.96\%)& 0& 267\\\hline
13 year& 99(6.97\%)& 0& 246\\\hline
14 year& 57(3.91\%)& 0& 228\\\hline
15 year& 71(4.87\%)& 0& 213\\\hline
16 year& 157(10.77\%)& 0& 204\\
\bottomrule
\end{tabular}
}}
\caption{\small Users in CTUs and BU groups with timelines of identical duration.}
\label{tab:total_activity_span}
\end{table}
%
\subsection{Consistently toxic users (CTUs) generate highly toxic content despite lower annual tweet volumes compared to BUs.}
\label{sec:comparison}

In this section, we look into our approach for understanding the time spans of activity exhibited by two distinct user groups on our platform, namely Consistently Toxic Users (CTUs) and Basic Users (BUs). Our objective is not only to delineate the temporal engagement of these groups but also to explore the potential implications of their behavior over time.

\begin{figure}[h]
    \centering
    \begin{subfigure}[b]{.32\textwidth}
    \centering
    \includegraphics[width=\textwidth]{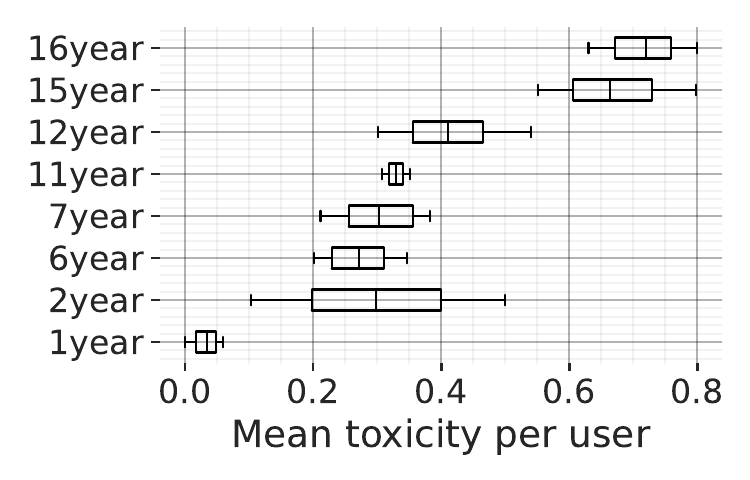}
    \caption{}\label{fig:toxicity_boxplot_per_year.pdf}
    \end{subfigure}
    
    \begin{subfigure}[b]{.32\textwidth}
    \centering
    \includegraphics[width=\textwidth]{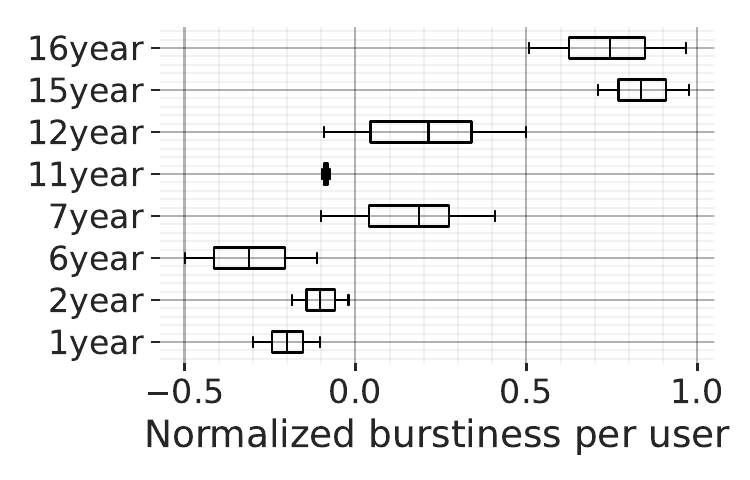}
    \caption{}\label{fig:burstiness_boxplot_per_year.pdf}
    \end{subfigure}
    %
    
     \caption{\small (\subref{fig:toxicity_boxplot_per_year.pdf}) Mean toxicity scores of users with different spans of activity; (\subref{fig:burstiness_boxplot_per_year.pdf}) Burstiness (all tweets) scores of users with different spans of activity. 
     }
    \label{}
\end{figure}

We utilize the approach described in Section~\ref{sec:Activity span} to ascertain the time spans of activity for both CTUs and BUs. This involves a detailed examination of the yearly timeframes during which these user groups actively participate on the platform. Through this method, we derive comprehensive insights into the duration of engagement for CTUs and BUs, allowing us to comprehend the evolution of their involvement over time.

The findings presented in Figure~\ref{fig:toxicity_boxplot_per_year.pdf} and Figure~\ref{fig:burstiness_boxplot_per_year.pdf} accentuate a noteworthy trend. CTUs, who consistently sustain their activity over extended periods, exhibit a higher level of toxicity and a more pronounced bursty behavior in their interactions. This phenomenon is complemented by a noticeably reduced annual tweet frequency (as discussed in Section~\ref{sec:Activity span annual}).

The implications of this intriguing pattern are profound. Although CTUs post fewer tweets within a given year, the content they share tends to be notably toxic. This suggests that a subset of long-term users may be responsible for a significant portion of harmful content on the platform. Additionally, these users showcase intermittent phases of inactivity between their active bursts, thereby underscoring the distinctive bursty posting pattern that characterizes their behavior. This raises questions about the platform's moderation policies and the effectiveness of addressing toxic behavior among long-term users.

\begin{figure}[t]
    \centering
    %
    \begin{subfigure}[b]{.32\textwidth}
    \centering
    \includegraphics[width=\textwidth]{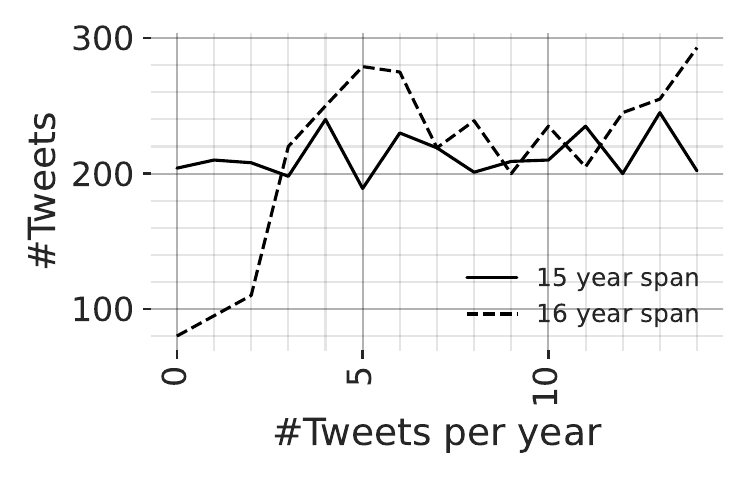}
    \caption{}\label{fig:tweets_15_16_example.pdf}
    \end{subfigure}
    
    \begin{subfigure}[b]{.32\textwidth}
    \centering
    \includegraphics[width=\textwidth]{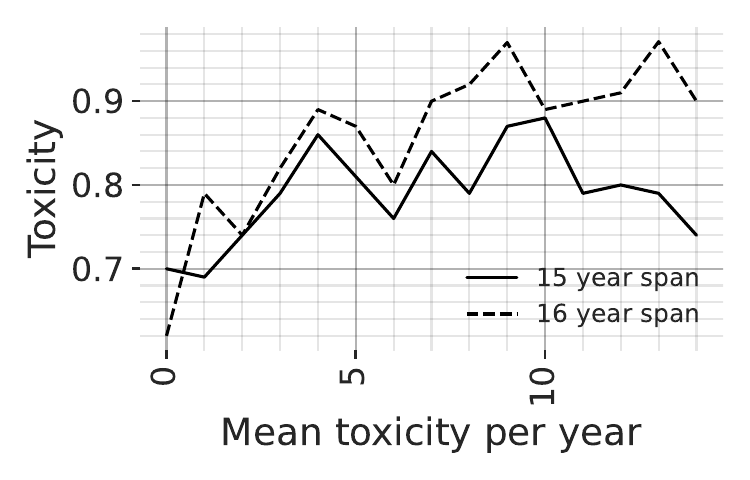}
    \caption{}\label{fig:toxicity_15_16_example.pdf}
    \end{subfigure}
    
     
     \caption{\small(\subref{fig:tweets_15_16_example.pdf}) Number of tweets of two users with 15 and 16 years of activity.; and (\subref{fig:toxicity_15_16_example.pdf}) Change in toxicity of the tweets of two users with 15 and 16 years of activity.}
    \label{}
\end{figure}

Illustrative expample: An noteworthy example emerges from the analysis of Figure~\ref{fig:tweets_15_16_example.pdf} and Figure~\ref{fig:toxicity_15_16_example.pdf}. These visuals distinctively showcase a connection between the escalation in tweet frequency by toxic users over time and a simultaneous increase in their level of toxicity. Of particular note is the consistent pattern of posting toxic content maintained by these users. This example underscores the dynamic relationship between posting frequency, toxicity, and temporal behavior.

\begin{figure}[!ht]
    \centering
    \begin{subfigure}[b]{.80\columnwidth}
    \centering
    \includegraphics[width=\textwidth]{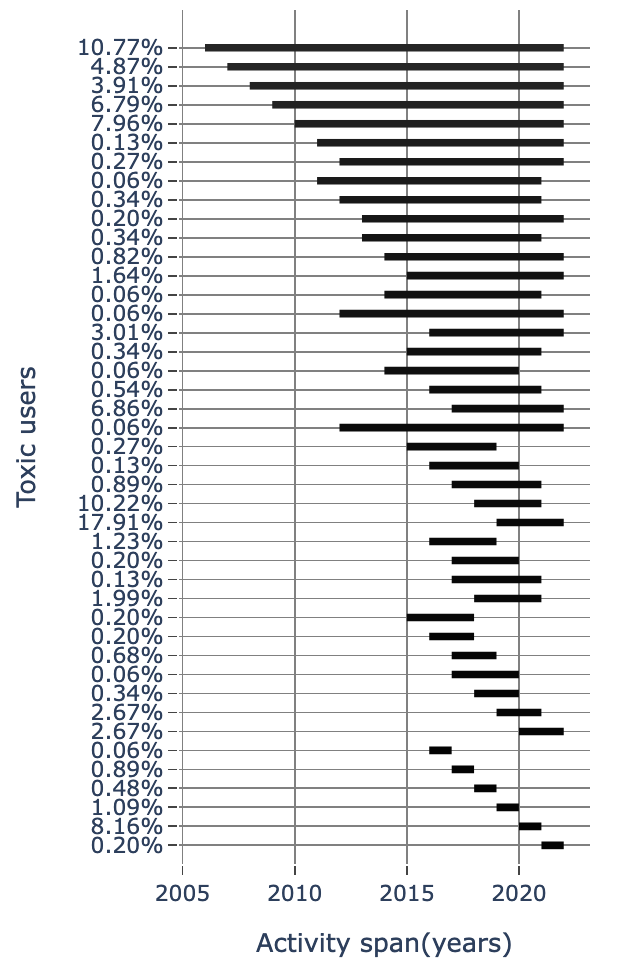}
    \caption{}\label{fig:newplot.png}
    \end{subfigure}
    
    \begin{subfigure}[b]{.60\columnwidth}
    \centering
    \includegraphics[width=\textwidth]{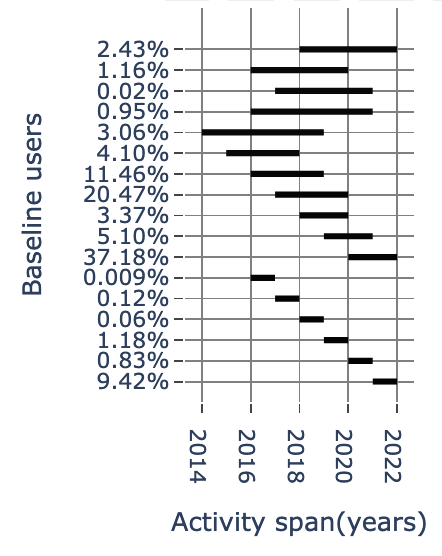}
    \caption{}\label{fig:newplot1.png}
    \end{subfigure}

     \caption{\small (\subref{fig:newplot.png}) Gantt chart illustrating yearly tweeting activity for CTUs; (\subref{fig:newplot1.png}) Gantt chart illustrating yearly tweeting activity of BUs.}
    \label{}
    \vspace{-5mm}
\end{figure}

Furthermore, we generate Gantt charts illustrated in Figure~\ref{fig:newplot.png} and Figure~\ref{fig:newplot1.png} to provide a more comprehensive visualization of users with varying activity span durations. These visualizations offer insights into how consistently toxic users (CTUs) exhibit consecutive years of activity and extended lifespans on the platform.

Understanding the temporal dynamics and behavioral patterns of user groups is essential for platform administrators and policymakers. It highlights the need for proactive moderation strategies that can effectively target long-term users with a propensity for toxic interactions. Moreover, it prompts a reevaluation of engagement metrics that consider not only quantity but also the quality and impact of user activity.

In conclusion, our in-depth analysis not only underscores the intricate relationship between user activity span, toxicity, and burstiness but also illuminates the need for a nuanced understanding of these patterns in the realm of online behavior. This newfound knowledge provides a robust foundation for more focused research initiatives and strategic interventions, all aimed at cultivating healthier online communities and elevating the overall user experience on the platform, paving the way for a more constructive and safer digital landscape.
\subsection{CTUs display shorter, more toxic bursts of behavior, while BUs engage consistently with lower toxicity over extended periods.}
\label{sec:temporal_analysis}

Twitter users' timelines are a dynamic canvas of time-dependent interactions, offering a valuable window into their online behavior. In this section, we delve into a meticulous exploration of these timelines, segmenting them into meaningful intervals. Our analysis is anchored in the framework of ``Churn analysis'' (as discussed in Section~\ref{sec:Churn analysis}), which sheds light on the ebb and flow of user activity.

To gain a deeper understanding, we partition the timelines of both Consistently Toxic Users (CTUs) and Basic Users (BUs) into weekly units. This segmentation reveals two distinctive phases: a ``life period" marked by continuous activity and a ``death period" characterized by consecutive weeks of inactivity. Within this analysis, we document cumulative toxicity levels and tweet counts per life period, peering into the prevalence of these life and death periods across users' entire timelines.

Previously, we classified each tweet from a Twitter user as either toxic (with toxicity scores surpassing the mean of their 3,200 tweets) or non-toxic (with scores falling below the mean), as elaborated in Section~\ref{sec:Consistently toxic users}. The consistency of toxic behavior hinges on two key metrics: average lifetime and average death duration. A substantial average lifetime signifies a week predominantly marked by toxic tweets, leading us to categorize users as ``Consistently toxic''. Conversely, a low average death duration underscores the resiliency of consistently toxic users, as they engage in non-toxic tweeting for a shorter duration.

\begin{figure}[t]
    \centering
    \begin{subfigure}[b]{.49\columnwidth}
    \centering
    \includegraphics[width=\textwidth]{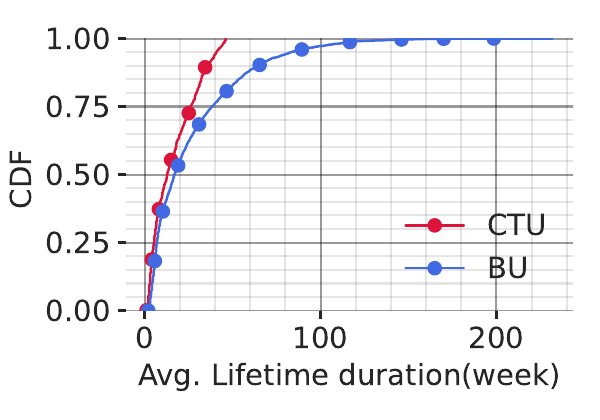}
    \caption{}\label{fig:churn_average_lifetime.pdf}
    \end{subfigure}
    \begin{subfigure}[b]{.49\columnwidth}
    \centering
    \includegraphics[width=\textwidth]{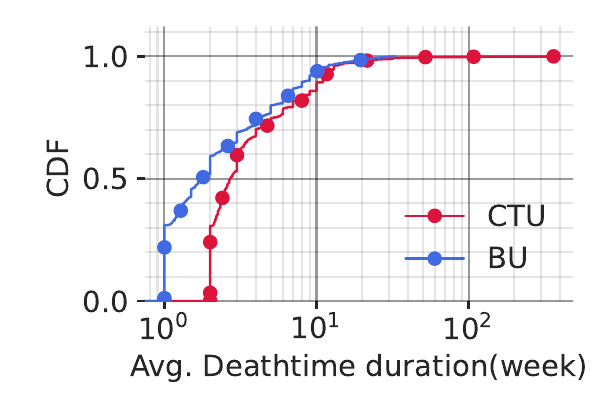}
    \caption{}\label{fig:churn_average_deathtime.pdf}
    \end{subfigure}
    
    \begin{subfigure}[b]{.49\columnwidth}
    \centering
    \includegraphics[width=\textwidth]{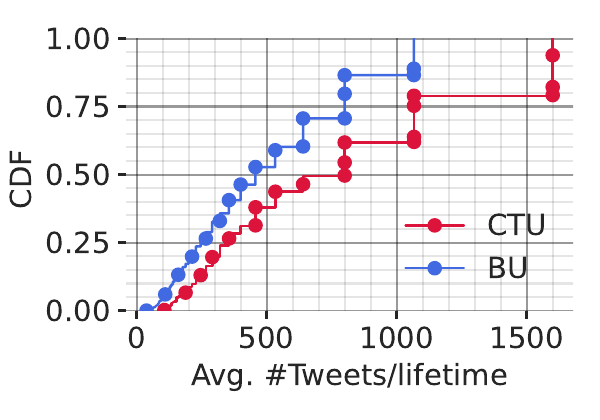}
    \caption{}\label{fig:churn_average_tweets_per_lifetime.pdf}
    \end{subfigure}
    \begin{subfigure}[b]{.49\columnwidth}
    \centering
    \includegraphics[width=\textwidth]{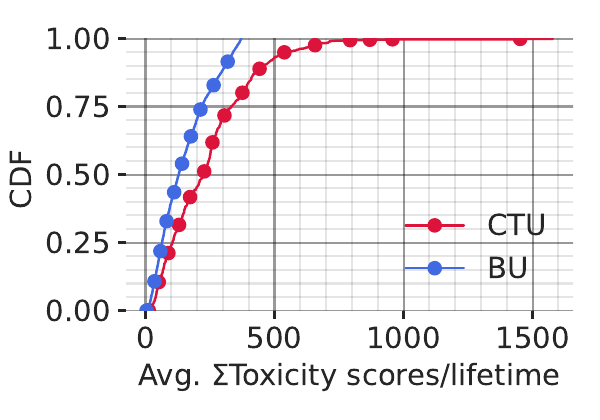}
    \caption{}\label{fig:churn_average_toxicity_lifetime.pdf}
    \end{subfigure}
    
    \begin{subfigure}[b]{.49\columnwidth}
    \centering
    \includegraphics[width=\textwidth]{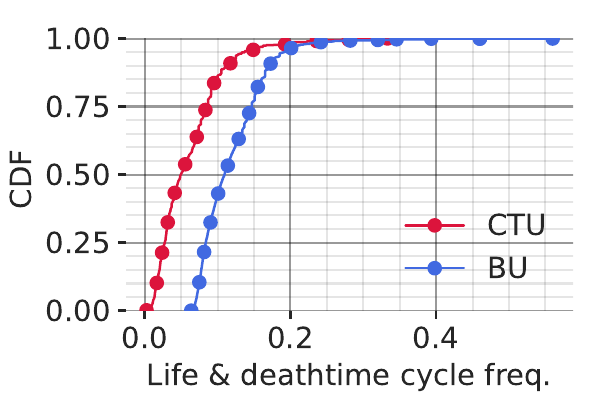}
    \caption{}\label{fig:lifetime_deathtime_lifetime.pdf}
    \end{subfigure}
    
    \vspace{-2mm}
     \caption{(\small \subref{fig:churn_average_lifetime.pdf}) An average lifetime of users in weeks toxic and benign groups(weeks); (\subref{fig:churn_average_deathtime.pdf}) Average deathtime of users in toxic and benign groups(weeks); (\subref{fig:churn_average_tweets_per_lifetime.pdf}) An average \#Tweets per lifetime(weeks); and (\subref{fig:churn_average_toxicity_lifetime.pdf}) An average $\sum$ toxicity scores per lifetime(weeks); (\subref{fig:lifetime_deathtime_lifetime.pdf}) Life and deathtime cycle frequency.}
    \label{fig:churn_analysis}
\end{figure}

Illustrated in Figure~\ref{fig:churn_average_lifetime.pdf}, a cumulative distribution function (CDF) unveils an intriguing pattern: CTUs exhibit briefer lifespans, with around 50\% experiencing less than 10 weeks of life duration, while another 50\% fall within the 10-20 week range. This finding has implications for understanding the volatility of toxic behavior among CTUs.

Figure~\ref{fig:churn_average_deathtime.pdf} showcases the mean duration of death periods, where BUs display relatively shorter death periods compared to CTUs. This aligns with our earlier observations and hints at the persistence of toxic users.

Figures~\ref{fig:churn_average_tweets_per_lifetime.pdf} and~\ref{fig:churn_average_toxicity_lifetime.pdf} reveal that CTUs' lifetimes involve more activity and heightened toxic intensity. These figures underscore the complexity of CTUs' behavior, characterized by extended periods of intense activity and toxicity.

Lastly, Figure~\ref{fig:lifetime_deathtime_lifetime.pdf} underscores a noteworthy difference: BUs exhibit more frequent life and death cycles compared to CTUs, highlighting the dynamic nature of user engagement among BUs.

In conclusion, our analysis unveils noteworthy patterns in the temporal engagement and toxicity dynamics of Twitter users. These findings not only deepen our understanding of online behavior but also hold implications for platform management and the development of more targeted moderation strategies to foster healthier online communities.

\section{Ethical Considerations}
\label{sec:Ethical Considerations}
The research detailed in this paper adheres to non-commercial principles, aligning with Twitter's Terms and Conditions for research endeavors. Throughout the course of our experiments, we meticulously observed ethical guidelines as stipulated in~\cite{rivers2014ethical}. In recognition of our experimentation involving data generated by human interactions, we sought and obtained formal approval from our institution's Institutional Review Board (IRB) under the reference Macquarie University IRB Project Reference: 35379, Project ID: 10008, Granted: 27/11/2021. It is important to underscore that our data will not be utilized for any commercial purposes, respecting the parameters of our research commitment.
\section{Discussion and conclusion}
\label{sec:discussion and conclusion}

The research presented in this paper represents a comprehensive exploration into consistently toxic users (CTUs) within the Twitter ecosystem, delving deep into the complexity of their behavioral patterns. The insights gleaned from this study carry significant implications, particularly in the domains of content moderation, online behavior research, and the overarching objective of fostering more secure and constructive digital environments. The findings unearthed herein serve as a pivotal stepping stone toward advancing our understanding of toxic behavior and enhancing the overall well-being of online communities.

One of the most remarkable findings is the revelation that CTUs exhibit strikingly human-like behavior online, as evidenced by their significantly lower Botometer scores compared to baseline users (BUs). This challenges the prevailing notion that automated bots are the primary culprits behind toxic interactions. Instead, it underscores the necessity of evolving content moderation strategies to account for the substantial contribution of CTUs, who closely mimic typical human engagement. Going forward, research should focus on refining tools and strategies capable of effectively distinguishing between toxic and benign human users, expanding the scope of moderation beyond automated bots and specific keywords.

Our analysis has illuminated the bursty tweeting patterns distinctive to CTUs, setting them apart from the more random posting behavior of BUs. Recognizing burstiness as a potential marker for toxic behavior is paramount for the development of content moderation strategies. Future research endeavors should center around the creation of advanced machine learning models and algorithms capable of detecting and responding to bursts of toxic content in real-time. This approach will effectively mitigate the impact of toxic interactions on the platform, fostering a safer and more constructive online community.

The revelation of a reduced annual tweeting rate among CTUs highlights the persistent nature of toxicity within certain online user profiles. This aspect opens an intriguing avenue for further exploration, aiming to decipher the underlying factors contributing to prolonged toxic engagement. Future research should delve into the psychological, sociological, and contextual elements that drive users to consistently exhibit toxic behavior over extended periods. Insights from such investigations can inform the development of targeted interventions aimed at mitigating long-term toxicity and fostering positive online interactions.

Another pivotal finding is that CTUs generate notably toxic content despite posting fewer tweets annually. This insight implies that a subset of long-term users may be responsible for a significant portion of harmful content on the platform. Consequently, future research should prioritize identifying and addressing consistently toxic users in content moderation strategies. These strategies should emphasize the quality and impact of user activity rather than merely assessing quantity.

Our temporal analysis has underscored the distinct engagement and toxicity dynamics between CTUs and BUs. CTUs exhibit briefer lifespans, more intense bursty behavior, and higher toxicity levels during their active periods, while BUs engage more consistently with lower toxicity over extended durations. This finding calls for nuanced strategies to tackle the volatile nature of toxic behavior among CTUs and the persistent nature of toxicity in the long term. Future research should explore the development of personalized interventions and support mechanisms tailored to the unique behavioral profiles of these user groups.

Beyond the confines of Twitter, the salient feature of our approach resides in its adaptability, signifying its intrinsic potential to be extended for application across various social media platforms. This inherent versatility equips researchers and practitioners with a potent instrument for the identification of consistently toxic users, rooted in their posting behavior, thus surmounting the platform-specific constraints.

In summary, this study sheds light on the multifaceted nature of toxic behavior on Twitter, emphasizing the need for comprehensive and evolving content moderation strategies. Future research should adopt interdisciplinary approaches that combine behavioral analysis, machine learning, psychology, and sociology to gain a holistic understanding of online toxicity. By continually refining our comprehension and interventions, we can pave the way for healthier, safer, and more constructive digital communities. Ultimately, the aim is not only to combat toxicity but also to create an online environment where users can engage in meaningful and respectful interactions, ensuring the well-being and integrity of digital spaces.

Looking ahead, several promising avenues emerge for future research. These include developing advanced machine learning models for real-time burst detection and content moderation, investigating the long-term effects of toxic behavior on the mental health and well-being of both users and online communities, exploring the role of platform design and user interface in influencing online behavior and toxicity, and collaborating with social media platforms to implement and evaluate proactive interventions based on the findings of this research. Embracing these future directions will enable researchers to contribute to the creation of a safer, more inclusive, and more positive online ecosystem for all users.
\bibliographystyle{plain}


\end{document}